**Letter to the Editor: Wall slip in dispersion rheometry.**

Richard Buscall, MSACT Consulting, Exeter, U.K.. r.buscall@physics.org.

*Whereas our understanding of the role and mechanism of wall slip has improved substantially over the last decade or two, it is still common to see papers on disperse systems appear wherein scant details of the measurement methods are given and where no mention of the possibility (probability?) of slip is made. It is argued that there is a need to raise awareness. It takes experience, judgement and skill to make meaningful measurements on disperse systems and it is suspected that the nature of the experimental challenge is under-estimated grossly by too many workers even now.*

That disperse systems tend to slip at smooth bounding surfaces, e.g. the surfaces of polished rheometer tools, has been known since the early days of rheometry (Mooney 1931). It is now known that a wide variety of disperse systems show wall slip, including, particulate suspensions [Buscall *et al*. (1993), Persello *et al.* (1994), Aral and Kalyon (1994) Barnes (1995), Pignon *et al.* (1996), Russel and Grant (2000), Walls (2003)], microgel dispersions [Cloitre *et al*. (2000), Meeker *et al*. (2004a,b), Seth *et al.* (2006, 2009)], emulsions and foams [Princen (1985), Plucinski *et al*. (1998), Pal (2000), Denkov *et al*. (2005)], worm-like micelles [Hu *et al.* (2002)] and vesicles [Cantat and Misrah (1999)]. Indeed, it should be well-known, arguably, since the review by Barnes (1995) is, to date, the most cited article to have been published in *J. non-Newtonian Fluid Mech.*. Furthermore, there appears to have been an upsurge of interest in the phenomenon of slip itself, as might be judged from the references cited here, of which ca. half those on slip are 21$^{st}$ C. In spite of this, it is common still to see many papers which make no mention of the possibility of slip and where the data have be generated using unsuitable or unspecified tools and methods. Furthermore, such papers continue to appear regularly even in specialist rheology and colloid science journals. This suggests not just that there is a lack of awareness amongst a sub-set of researchers and authors, but also that the same applies to some referees. In the case of *Journal of Rheology* (JOR), this is perhaps somewhat ironic, given that papers concerned with the phenomenon of slip itself have appeared regularly over the last two decades, as again can be judged from the list papers cited. Thus, and by way of example, whereas last year saw the publication in JOR of an important paper on the mechanism of wall-slip by Seth *et al*. (2008), quarter-one 2009 has seen three papers appear concerning particulate systems of which only one addresses or acknowledges the possibility of the slip [Carrier and Petekides (2009)]. In a paper by Grillet *et al*. (2009) on thixotropic dispersions of a type known to slip and a paper by Haleem and Nott (2009) on filled polymer solutions, there appears to have been no consideration given to the matter. The concern here is not just that slip can cause the viscosity (etc.) to be in serious error, nor just that slip can give rise to false thixotropy [Buscall et al. (1993), Barnes (1995)], it is also that slip is much more than a nuisance or a source or error, it is fundamental to the way disperse systems respond and behave.

That disperse systems show slip should not be too surprising perhaps, since, unless the bulk structure is maintained right up to and is coupled to the wall, and unless the disperse and continuous phases move exactly in tandem on all length scales $\lambda$, including $\lambda/R <1$ (where $R$ is particle size), then slip will have occurred to some extent, in effect. Now, whether or not the slip is of any consequence in a rheometrical experiment is another matter. In the case of viscosity measurements, for example, it depends in part upon the ratio of the apparent slip-layer thickness $\delta$ (viz, the effective thickness of a notional thin film of pure solvent), to the gap width $g$ and in part upon the relative viscosity of the system. Whence it can be shown trivially that slip will not make a noticeable difference unless,

$$\frac{\eta}{\mu}\frac{\delta}{g} \sim \frac{\eta}{\mu}\frac{R}{g} > e \qquad (1$$

Where $\eta$ is the dispersion viscosity, $\mu$ that of the solvent (or solution, if the medium happens to be one) and $e$ is a discernable or unacceptable fractional error value. Thus, by way of example, if the scaled slip layer thickness, $\delta^* = \delta/g \sim 10^{-4}$, which is not untypical of a colloidal system, then noticeable errors will develop if the relative viscosity is 100 or more. To be more precise, the error is given by,

$$e = \frac{x}{1+x}; \quad x = \frac{\eta}{\mu}\frac{\delta}{g} \qquad (2$$

from which it can be seen that the error can be arbitrarily large if the relative viscosity is large; errors of 99 % or more have been demonstrated [e.g. Buscall 1993]. The apparent slip layer thicknesses determined by Buscall et al. (1993) for depletion-flocculated latex are replotted in fig.1. They are plotted against (0.59 – φ), where φ is the volume-fraction and 0.59 is that at the hard-sphere glass transition. Several points emerge: Notice firstly that at lower concentrations, δ can be several particles diameters, whereas it becomes very small indeed at high densities. Very small does not mean "negligible" though, far from it, since the viscosity error can still be 99% or more. Notice, secondly, that the slope of the trend line is ca. 2.5. According to the model of Seth et al. (2009), the slip layer thickness is predicted to vary with the shear modulus of the particulate gel $G$ like $G^{-2/3}$, everything else being equal. Note that in their model, Seth et al., used $1/G$ as a proxy for the compressibility of the particulate phase. The exponent of 2.5 from fig. 1 then, should correspond to concentration exponents for the modulus and compressive strength of ca. 3.7 and whereas the modulus and compressive strength have not been reported for the particular depletion system used by Buscall et al. (1993), many such measurements have been made for many other weakly and strongly flocculated systems and an exponent of 3.5 – 4, is found to be typical, as can be seen from the small sub-set of sample data re-plotted in fig. 2.

The problem of slip can be avoided very easily of course by using milled, serrated, or otherwised roughened or textured tools. The amplitude of the texture need not be all that large perhaps. Thus Buscall et al. (1993) found that sand-blasting was sufficient to eliminate slip with sub-micron particles (even though it is not for larger particles). In an elegant experiment Isa et al. (2007) demonstrated *inter alia* that glueing a monolayer of the particles themselves onto the wall of a capillary was suffficient to eliminate slip. It would appear then that the texture need only be on the scale of the particle size (even though erring on the side of caution would seem prudent). Most rheometer manufacturers will now supply a selection of roughened and serrated tools, albeit not necessarily as standard or by default, which may be part of the problem, perhaps. There may a question of mind-set also: The author once heard somebody accused of neglecting slip say "Yes, I am aware of the possibility of slip but did not want to compromise the accuracy with which the gap was set and defined". Now, one can be sure that this person did not actually mean "I am prepared to risk the possibility of an error of unknown and therefore arbitrarily large magnitude in order avoid an uncertainty in gap setting of perhaps 1%", except that is what it amounts to.

It has been the experience of the writer and his colleagues over ca. thirty-five years of testing a very wide range dispersion and emulsions that most disperse systems show slip. There do seem to be exceptions, some colloidally-stable dispersions of genuine nanoparticles ($R << 100$ nm) seem not to, although this may simply be that even the optical or sub-optical scale of roughness of ordinary rheometer tools, e.g. those made of anodized aluminium as used in the writer's own work on such systems Buscall et al. 1982), suffices to negate slip when the particle size is *very* small. It has been the experience of the writer also that whereas some aggregated or flocculated dispersions and emulsions can and do show true thixotropy, it is more usually encountered when the solvent is viscous. Thus, dispersions of the type used by Grillet *et al* (2009), that is, dispersions of metal oxide particles in real or model resins, are often found to be genuinely thixotropic. That is not to say though that the thixotropy is not modified and exacerbated by slip when such systems are tested with smooth tools, it is.

The onset of shear-thinning can be associated with some characteristic shear stress value, $\sigma_c$, say, which then becomes the yield stress $\sigma_y$ in the limit of arbitrarily large low-shear viscosity. It has been found also that there can be a characteristic or critical stress associated with slip [Buscall *et al*. (1993), Barnes (1995), Tindley (2007), Seth et al (2008)]. In the case of colloidally-stable systems this stress can be very small [Seth et al (2008)], relatively, speaking, whereas with flocculated systems the wall critical stress can be closer in magnitude to the bulk characteristic stress, as a result of adhesion to the walls [Buscall *et al*. (1993, Tindley (2007)]. Now, this raises an interesting point, given that, everything else being equal, the force of attraction between a particle and a wall is predicted to be twice that between two particles. This might be taken to imply that the wall stress need no necessarily be less that the true yield stress in all cases; perhaps it depends upon the materials of construction [Seth *et al.* (2008)]. There is a second interesting point associated with the vane method of determining true yield stresses [e.g. review by Barnes and Nguyen (2001)]. In this method a 4-bladed vane (typically) of diameter *d* is immersed in a sample confined in a wide cylinder of diameter *D* and, more often than not, the cylinder has smooth interior walls. A wide gap (*D-d*)/*d* is used to avoid premature yield at the walls as a result of slip, notice though that because the stress decays radially this method relies on the inequality,

$$\frac{\sigma_w}{\sigma_y} > \left(\frac{d}{D}\right)^2 \qquad (3$$

since, otherwise, wall-slip would still occur. That it is possible to switch at will from bulk to wall yield by varying D has been demonstrated in detail by Tindley (2007).

The existence of a critical wall stress might be taken to imply that linear viscoelastic measurements can safely be made using smooth or polished tools and whereas, this can be done, care has to be taken to ensure that the microstructure has been given sufficient time to attach or adhere to the wall before the measurement is started. Thus, Buscall *et al.* (1993) found that depletion-flocculated dispersions of sub-micron particles took an hour or more to bond to stainless steel tools, even though bulk thixotropic recovery was instantaneous in effect.

An alternative but well-established method of characterising materials that slip is to use smooth tools but to vary the gap [Yoshimura and Prud'homme (1988)]. This method is attractive in principle, although in practice it can have its limitations since, often, the gap cannot be varied by all that much for one reason or another (e.g. ejection of the sample or fracture in the case of parallel discs). An alternative way to proceed when, say, the need is to understand how to scale data in the context of an application involving smooth surfaces, is to test with both rough and smooth tools [Buscall(1983)].

Whereas the proportion of papers failing to acknowledge slip seems to have reduced slowly the last decade or so, a concern is that it may be approaching a steady-state, i.e. that there could well be parts of the community where a lack of awareness and a lack of appreciation of the very real challenges and difficulties of experimental dispersion rheology is passing from one generation of researchers to the next. Editors and referees might need to be more vigilant perhaps, even though, in the experience of the writer it can be very difficult to persuade authors to address the problem or possibility of slip after the event; it is then too late. This would imply then that the solution lies more with rheometer manufacturers, trainers and educators perhaps.

Should the motto of disperse-system rheology sub-section be "Everything Slips"? Well, perhaps not, but it might make a good working principle, even if not quite everything does. It would certainly be better than embarking upon dispersion rheometry with the counter-view, or in a state of blissful ignorance.

**Figures: - next two pages**

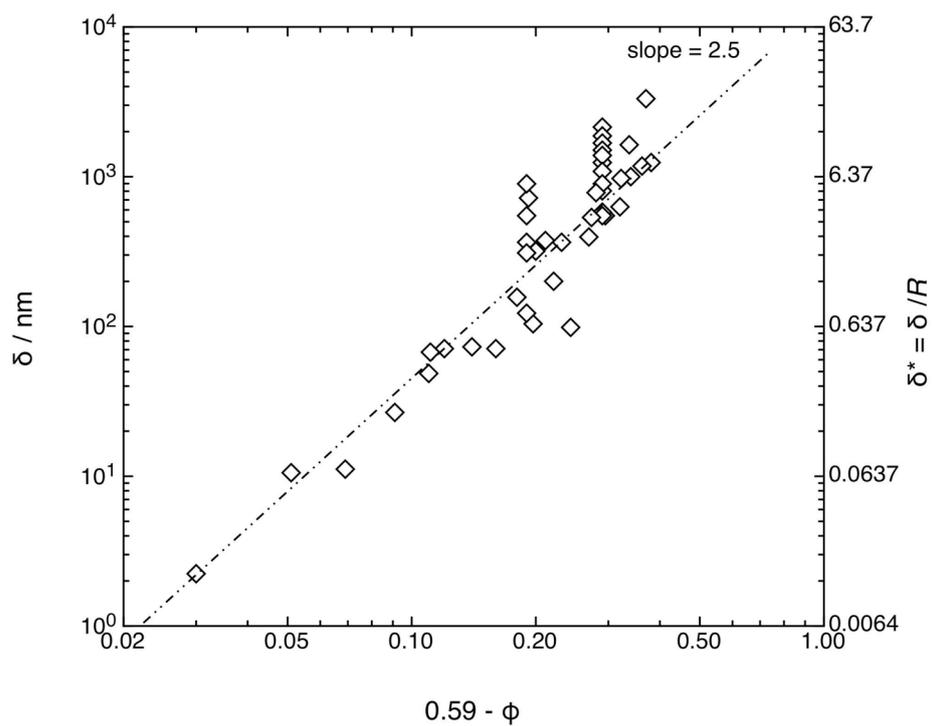

**Fig.1** Apparent slip layer thickness δ (left) and scaled thickness δ/$R$ (right) versus volume-fraction for depletion flocculated latex of mean particle radius $R$ = 157nm. Data taken from Buscall *et al*. (1993). A concentration exponent of ca. 2.5 is indicated.

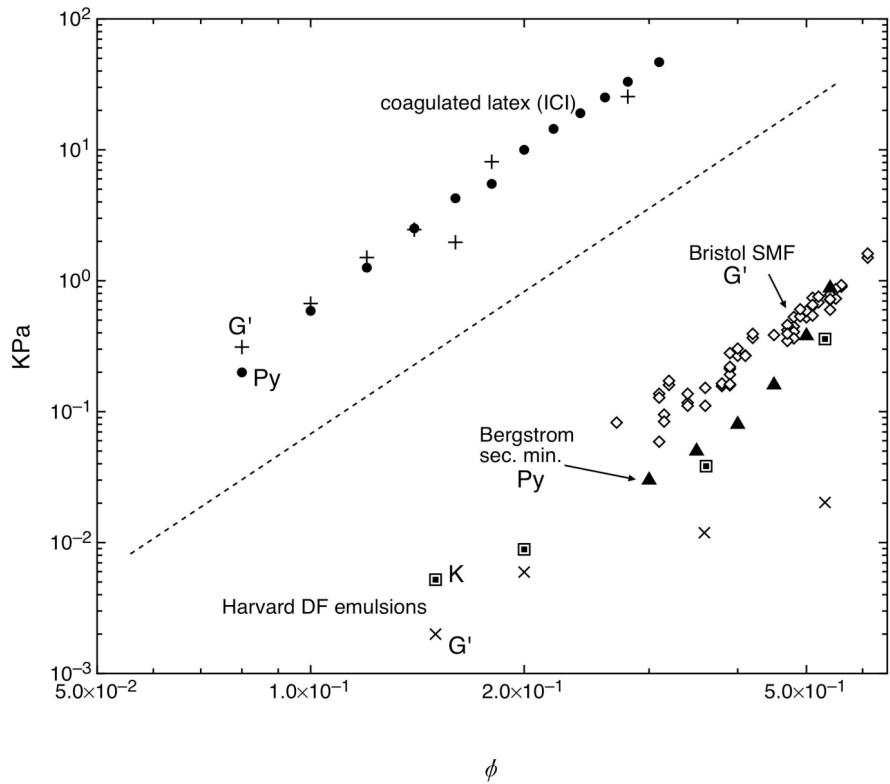

**Fig. 2** Some examples of shear modulus *G* and compressive strength $P_y$ plotted against volume-fraction for both weakly and flocculated dispersions and emulsions. Crosses + shear modulus data of Buscall *et al.* (1988) for coagulated PSL, filled circles - $P_y$. Diamonds – A similar latex weakly-flocculated [Partridge (1985)]. Filled triangles – the compressive-strength of secondary-minimum (SMF) flocculated alumina [Bergstrom *et al.* (1992)]. Crosses x – *G* for depletion flocculated emulsions [Kim et al. (2007)], semi-filled squares, compressional modulus, $K= dP_y/d\ln\varphi$ for the same. Note that in this fig. the abscissa is φ, not 0.59-φ, as in fig.1. The reason for this is that *G* is not expected to diverge at the glass transition whereas $P_y$ and *K* are. It can be seen though that where the latter data extend to very high concentration, as do those for the weakly-flocculated systems of Bergstrom *et al.* and Kim *et al.*, upwards curvature is evident. The drawn line has a slope of 3.75, this being the slope that might be expected from fig. 1 in the light of the model of Seth *et al.* (2009).

10